\documentclass[MSNbibl,nameyear,dvips]{arxstspdf}
\usepackage{flushend}
\usepackage{stfloats}
\volume{29}
\issue{1}
\pubyear{2014}
\firstpage{103}
\lastpage{103}
\doi{10.1214/14-STS466} % kopijuoti is 'New paper accepted'
\referstodoi{10.1214/12-STS413}% pagrindinio straipsnio DOI, kai

\begin{document}
\begin{frontmatter}
\vspace*{6pt}
\title{Reply to the Discussion of ``Estimating the Distribution of
Dietary Consumption Patterns''}%\thanksref{T1}
% kai straipsnis turi susijusiu diskusiju ir rejoinder'iu
%rejoinder at \relateddoi{r}{10.1214/00-STSXXXX}.}
\runtitle{Dietary Consumption Patterns}

\begin{aug}
\author[a]{\fnms{Raymond J.} \snm{Carroll}\corref{}\ead[label=e1]{carroll@stat.tamu.edu}}
\runauthor{R. J. Carroll}

\affiliation{Texas A\&M University}

\address[a]{Raymond J. Carroll is Jill and Stuart A. Harlin Chair
and Distinguished Professor of Statistics, Nutrition and Toxicology,
Department of Statistics, Texas A\&M University, 3143 TAMU, College Station,
Texas 77843-3143, USA \printead{e1}.}

\end{aug}

% ABSTRACT

% KEYWORDS

\end{frontmatter}

%s1 #&#

The discussants' repeat remarks they raised in the review of the
original paper upon which my article is based (Zhang et al., 2011b), so I will be brief. As
part of their review we also produced extensive Supplementary Material:
\begin{enumerate}[3.]
\item[1.] In a time when it is routine to see models with thousands of
latent variables and small sample sizes, it seems far-fetched at best
to call ours ``\textit{highly complex}.'' We have 6 latent variables,
a few regression parameters, two 19-dimensional covariance matrices and
typically hundreds of observations, and thousands in our particular
application. If that is ``\textit{highly complex},'' what is being
discussed in this issue must be worthy of international awards.

\item[2.] Our model is fully parametric, not ``\textit{semiparametric}.''

\item[3.]``\textit{How, without something like sensitivity analysis, are
we to know that it is valid}?'' As mentioned in my article, and the
original paper, unlike massive latent variable problems, our model can
be checked, because the simple bivariate submodels for every
combination of dietary components can be fit by other means using
standard weighted likelihoods, and we have confirmed that the bivariate
sub-fits from our model agree with the direct fits to the submodels.
See Section~4 of the \textit{Annals of Applied Statistics} article
where this issue is discussed in more detail.

\item[4.] We are sorry that the Bayesian sample survey methodology is a
``\textit{mess}.'' This is too bad, on many fronts. Our method would be
fully Bayesian if there were no sampling weights needed.

\item[5.] We had an important, practical problem to be solved. The
percentage of children with alarmingly
bad diets is grossly
overestimated by the standard single 24 hour recall. Should we have
waited for the ``\textit{mess}'' to be cleared up before solving it
and being pure Bayesians in the process? I would love to do it as fully
Bayesian, but one needs to remember that, as said in the article, it
was Bayesian thinking that enabled the model to be fit to begin with,
for example, the latent variables we use are a standard Bayesian
formulation, MCMC computation, etc.

\item[6.] Section~3.3 of Zhang et al. (2011b)
%the \textit{Annals of Applied Statistics}
discusses weighting. We showed that the weights were not important for
model fitting, and conjectured it was ``\textit{because the covariates
we use are major players in determining the sampling weights}.''

\item[7.]``\textit{pseudolikelihood}'': Yes, indeed, see the ``\textit
{mess}'' above, the abstract and Section~4 of my article, and Section~4
of the \textit{Annals of Applied Statistics} paper.

\end{enumerate}

\textit{An Offer}: Understanding the distributions of usual intake in
a population is important. Bayesian thinking allowed us to propose and
fit our model, the methodology is fully Bayesian in nonsurvey problems,
and is being used in such contexts to analyze the risk of usual dietary
intake on disease.

I would be happy to work with any expert in Bayesian survey methodology
to extend our work to be properly Bayesian in the context of NHANES,
with proper Bayesian uncertainty statements. This problem is not going
away, and regularly the analyses need to be updated.\vfill


%suskaldyti doi

\end{document}